# Di-nucleotide Entropy as a Measure of Genomic Sequence Functionality

Dmitri V. Parkhomchuk

Email address: parkhomc@molgen.mpg.de , pdmitri@hotmail.com

**Abstract:** Considering vast amounts of genomic sequences of mostly unknown functionality, *in-silico* prediction of functional regions is an important enterprise. Many genomic browsers employ GC content, which was observed to be elevated in gene-rich functional regions. This report shows that the entropy of di- and tri-nucleotides distributions provides a superior measure of genomic sequence functionality, and proposes an explanation on *why* the GC content must be elevated (closer to 50%) in functional regions. Regions with high entropy strongly co-localize with exons and provide genome-wide evidences of purifying selection acting on non-coding regions, such as decreased SNPs density. The observations suggest that functional non-coding regions are optimised for mutation load in a way, that transition mutations have less impact on functionality than transversions, leading to the decrease in transversions to transitions ratio in functional regions.

## Introduction

Entropy is a convenient and, in many respects, unique measure of variability of discreet variables [1, 2]. It is defined as

$$H(X) = -\sum_i P(x_i) \log(P(x_i)),$$

Where a random variable is distributed on the values $x_i$ with probabilities $P(x_i)$. For a given number of possible states this function has a maximum when the probabilities of all states are equal. The applications of entropy and information theory principles proved to be useful in providing the fundamental insights on genetic sequence evolution [3].

On the one hand, one of the interpretations of entropy is the information content of the message. On the other hand, it also can be interpreted as the randomness of the message. However the first interpretation is subjectively more appropriate for genomic sequences where we assume that entropy reflects the degree of functionality. The reasons for this are that the average GC content in a genome usually deviates from an ideal 50% (40% in human genome) lowering the entropy of unconstrained sequences, and such sequences undergo not randomisation by point mutations as could be expected, but homogenisation by repetitious sequences (which have low variability) performed by numerous processes like slippage replication, retroposition, etc. Local correlations in sequence composition are present genome-wide even in coding regions, indicating the presence of such homogenisation force. This force is opposed by "randomisation" force as the requirement for the most efficient coding in terms of information theory. For example the functionality of a sequence implies interaction specificity, thus the higher sequences variability (entropy) is, the large number of specific interactions such sequences may perform. Hence the assumption is that sequence entropy may serve as the averaged measure of functionality, aiding the localisation of non-coding functional regions. Considering DNA sequence as a message composed of 4 letters ACGT, the entropy may be computed for the distributions of single nucleotide or for the words consisting of few nucleotides. For one nucleotide, the entropy reflects the balance between 4 possible states – A, C, G, T, the maximum being when all their probabilities (computed by frequencies) are ¼, while for di-nucleotide words 16 variants are possible – AA, AC, etc.

Of course a region with low entropy cannot be immediately qualified as non-functional, because its functionality may be "sparse" - confined to a small sub-region surrounded by sequences of low complexity, thus its average functional load is low anyway. However the regions with exceptionally high entropy are most likely densely packed with functional elements. The provided analyses reveal several independent evidences to this information theoretical inference.

The question is what sequence length should be used for calculation because short sequences provide low statistics for di-nucleotides occurrences probabilities estimations (discretization) while large sequences lose position specificity to localise functional elements. So in performed analyses the human genome was split into pieces with lengths ranging from 5 kbp to 100 kbp (Fig. 1) and all of them reveal about the same robust trends.

## Results

### Co-localisation with exons

The human reference genomic sequence from NCBI release 35 was used. X and Y chromosomes were excluded because of their special mode of inheritance and thus genetic variability (i.e. SNPs distribution, as SNPs distribution analyses were used later). The total length of analysed sequence was $2.5 \cdot 10^9$ bp. This sequence was split in pieces of given size (ranging from 5 kbp to 100 kbp). For each piece the entropies were calculated for single nucleotide, di- and tri-nucleotide words - $H_1$, $H_2$ and $H_3$ correspondingly (Fig. 1). The pieces, which partially overlap with known exons, were treated separately.

The histograms of number of pieces with given entropy for different lengths are shown on Fig. 2. Entropy $H_1$ was not included because it turned out to be too crude as a metric of variability (the same with GC content) and shows no trends described. For example a simple repeat 'ACGT' would produce the maximum of $H_1$ entropy, while being apparently of low complexity. For $H_2$ and higher orders it is obviously more difficult to form the simple short repeat with maximum entropy.

Analysing Fig. 2 a number of interesting conclusions can be drawn. The right shoulder of distribution for exon-containing pieces is always above the distribution of no-exon pieces even for small 5 kbp pieces. Thus, for $H_2$ above the certain value for a given sequence chunk length we have more chances to find an exon inside that chunk than not to. Such co-localisation specificity cannot be achieved with GC content or $H_1$ measure.

The large 100 kbp pieces are also interesting in showing the apparent bimodality with rather abrupt transitions. At $H_2 > 3.93$ abrupt transition in exons density happens (Fig. 2).

**SNPs distribution**

Next analyses were concerned with SNPs densities. For this we used human SNPs from dbSNP release 125 (about $1 \cdot 10^7$ SNPs) and HapMap CEU [4] population ($\sim 3 \cdot 10^6$ SNPs). DbSNP provides the SNP average frequency only for a small part of entries most of which were derived from HapMap. DbSNP and HapMap are not independent as the SNP panel for HapMap was selected mainly as a subset of validated dbSNP entries, while HapMap provided back the frequencies of genotyped SNPs. In HapMap the frequency is given for a sample of 90 individuals (30 couples of unrelated and 30 their children).

Fig. 3 presents the density of SNPs versus entropy and thus supposedly the functionality.

The SNP density apparently drops at high $H_2$ for both dbSNP and HapMap. However the decrease in HapMap SNP's density is more dramatic – it's nearly twice fewer SNPs in high-H regions in comparison with low-H regions. The decrease starts at $H_2=3.93$ where the peak of exon-containing regions is located. DbSNP performs better because numerous rare SNPs are included in it. The number of rare SNPs in dbSNP is increased because the sequences around exons were deeply re-sequenced for numerous candidate genes in different ethnic groups. To support these inferences we performed the same analyses for a mouse SNPs that are supposed to have more homogeneous resequencing depth - most of the mouse SNPs were derived from the whole-genome sequencing with a small sample size. Mouse SNPs have the same abrupt decrease in density after $H_2 = 3.93$ and the same as for HapMap 50% decrease of SNPs density for genomic pieces with maximum entropy (Fig. 5). The initial increase in the SNP density (Fig. 5) is probably due to the increasing with entropy GC content and mutagenesis correspondingly. For HapMap (Fig. 3), this increase is weaker probably due to the strategy of SNPs selection for genotyping. Thus the HapMap SNPs panel and the mouse SNPs represent more frequency-homogeneous coverage (particularly in non-coding regions, as HapMap was focused to cover the coding regions) while human dbSNP SNPs are enriched in multiple rare SNPs, thus they are not reflecting the true amount of common sequence variability and functional constrains.

SNPs density drop remains the same if we discard exons from analyses, so the exons conservation cannot be used to explain it. In fact, even in the exon-containing pieces the length of exons is much smaller in average than the length of non-coding content: there are about 35,000 of 5 kb exon-overlapping pieces while the average gene length is about 1 kb. Thus the exons occupy in average less than 20% of exon-containing 5 kbp pieces. The coding sequence *per se* does not contribute significantly to the entropy and SNP density of large pieces of a DNA they are located in. Hence Fig. 2 suggests that exons are a minor part of the functional genomic sequence in large genomes. As a very rough estimate it is interesting to note that the average SNPs density in exons is about 70% from that in introns [6] while non-coding chunks with $H_2 > 3.94$ have in average about the same decreased SNPs density but few times (~80 Mbp) larger total length (than exons).

No significant dependencies of the SNPs frequencies on $H_2$ were observed, however a HapMap sample size might be too small to reveal them.

The SNPs density trends are easily reproducible and apparently highly significant because the relative error in density mean value is $\propto \frac{1}{\sqrt{N}}$ where $N$ is the number of chunks used for averaging in the corresponding entropy bin, which is quite large (Fig. 2). The magnitude of relative error is also apparent from point-to-point fluctuations (Fig. 3) as each point represents the average over independent genomic chunks.

**Ratio of transversions to transitions**

The most common mutations in a human genome are transitions (C-T, G-A), which are nearly twice more common than transversions. If we plot the ratio of transversions to transitions versus $H_2$ we observe the decrease in high-$H_2$ regions (Fig. 4). It is difficult to explain this by sequence composition because at high H the GC content is about 50% and is changing insignificantly in the region of 3.9-3.95. The proposed explanation for this is the action of purifying selection, which eliminates transversions as being more deleterious than transitions. Presumably, in average the non-coding functional sequences have consensuses, which are optimised for mutation load, and because most of the mutations are transitions, the sites evolved to accept transitions as "synonymous" i.e. fit in the consensus, as demonstrated by splicing sites [6] and the genetic code. The recovery of this ratio at the extreme values of $H_2$ (Fig. 4) can be explained by observation that the decrease of mutation load by transitions preference is possible only with semi-conserved sites [6] however at the extreme $H_2$ many sites are highly conserved.

**2D SNP distribution**

The same trends can be visualized on 2D plots. We plotted SNP densities with X coordinate being di-nucleotide entropy and Y – GC content. A number of interesting phenomena can be observed (Fig 6).

It is evident that the entropy and GC content provide rather independent information about sequence properties and they are not mutually redundant. The low-H high-GC "cloud" of dbSNP SNPs in the lower left corner seems to represent mutationally degraded GC-rich gene-poor repeats (such as ALU), with high SNP density. Probably these regions are intentionally properly underrepresented in HapMap. It is also evident the smaller density of HapMap SNP in high-H (starting at ~3.93 bits) regions which are exon-rich and have high load of non-coding functionality.

**Formal solution to C-value paradox?**

If the assumptions on coding efficiency (here 'coding' is an information theoretical term unrelated to amino acid coding) are correct, it can help to resolve some paradoxes of mapping the genetic complexity into phenotypic. For example *D. melanogaster* and *C. elegans* have approximately equal genome sizes and gene numbers, however a fly is perceived as much more complex phenotypically. On the plot of their genomic di-nucleotide entropy distributions (Fig. 7) it is clear that the fly genome has potential for far larger functional load, considering that functional load abruptly increases after $H_2 = 3.93$, which is in accord with their complexities. A fly has the compact genome presumably due to a high deletion rate [5], so it must utilise more efficient coding to fit more functional information into the same genomic length as more primitive nematode. Interestingly *D. melanogaster* entropy peak is located even at higher values (H~3.95, compare Fig. 7 and Fig. 2) than it is for human, probably reflecting high demands for its genome size compression, which are not present in large human genome.

**Conclusion**

Di-nucleotide (and higher orders) entropy seems to provide the measure of sequence functional load supported by the following theoretical and empirical arguments:
- Information-theoretical consideration, i.e.: multiple diverse functions require diverse sequence, while non-functional sequences tend to be homogenized to low complexity.
- Strong co-localisation of high-H non-coding sequences with exons.
- Decrease in transversions to transitions ratio in high-H regions. This evidence of functional load is yet in the stage of hypothesis itself though [6].
- Significant drop of the SNPs density in high-H regions, which is difficult to explain by sequence composition and/or mutagenesis. Thus the prime suspect is strong purifying selection in these regions.

This entropy measure may help to optimise the search for non-coding functional elements in the vast space of large genomes. For example if some 5 kb sequence has

$H_2 > 3.93$ it is likely densely functional even if there are no exons in it. The discovery and validation of non-coding elements can be further enhanced by a novel observation that such elements are optimised for mutation load by favouring transition mutations. Hence, the ratio of transversions to transitions is analogous to the replacement-to-synonymous (R/S) metric in the coding sequences. On the other hand the degenerate 3rd codon positions can be considered as non-coding sequences themselves, so for high-H genes they can be functionally overloaded challenging the validity of R/S metric in such genes.

While "C-value paradox" is not inasmuch the paradox any more, in the case of *C. elegans* and *D. melanogaster,* the information theory considerations provide coherent estimate of their relative complexities, which are difficult to match by their C- or G- values alone.

This metric opens the opportunities for numerous intra- and inter-genomic analyses providing the general estimate of functional load in non-coding sequences. It also suggests optimisation strategies for the target regions selection for resequencing and genotyping.

**References:**

1. Shannon, C. **A mathematical theory of communication.** *Bell System Technical Journal* **27**, 379-423 and 623-656 (1948).

2. Jean-Bernard Brissaud**, The meaning of entropy.** *Entropy* **7**[1], 68-96 (2005).

3. Schneider TD. **Evolution of biological information.** *Nucleic Acids Res.* 2000 Jul 15;28(14):2794-9.

4. Altshuler, D., Brooks, L.D., Chakravarti, A., Collins, F.S., Daly, M.J., Donnelly, P. **A haplotype map of the human genome. International HapMap Consortium.** *Nature* 437(7063),1299-320 (2005).

5. Petrov DA, Lozovskaya ER, Hartl DL. **High intrinsic rate of DNA loss in Drosophila.** *Nature.* 1996 Nov 28;384(6607):346-9.

6. Parkhomchuk DV. **Genetic Variability of Splicing Sites.** http://xxx.lanl.gov/abs/q-bio.GN/0611060

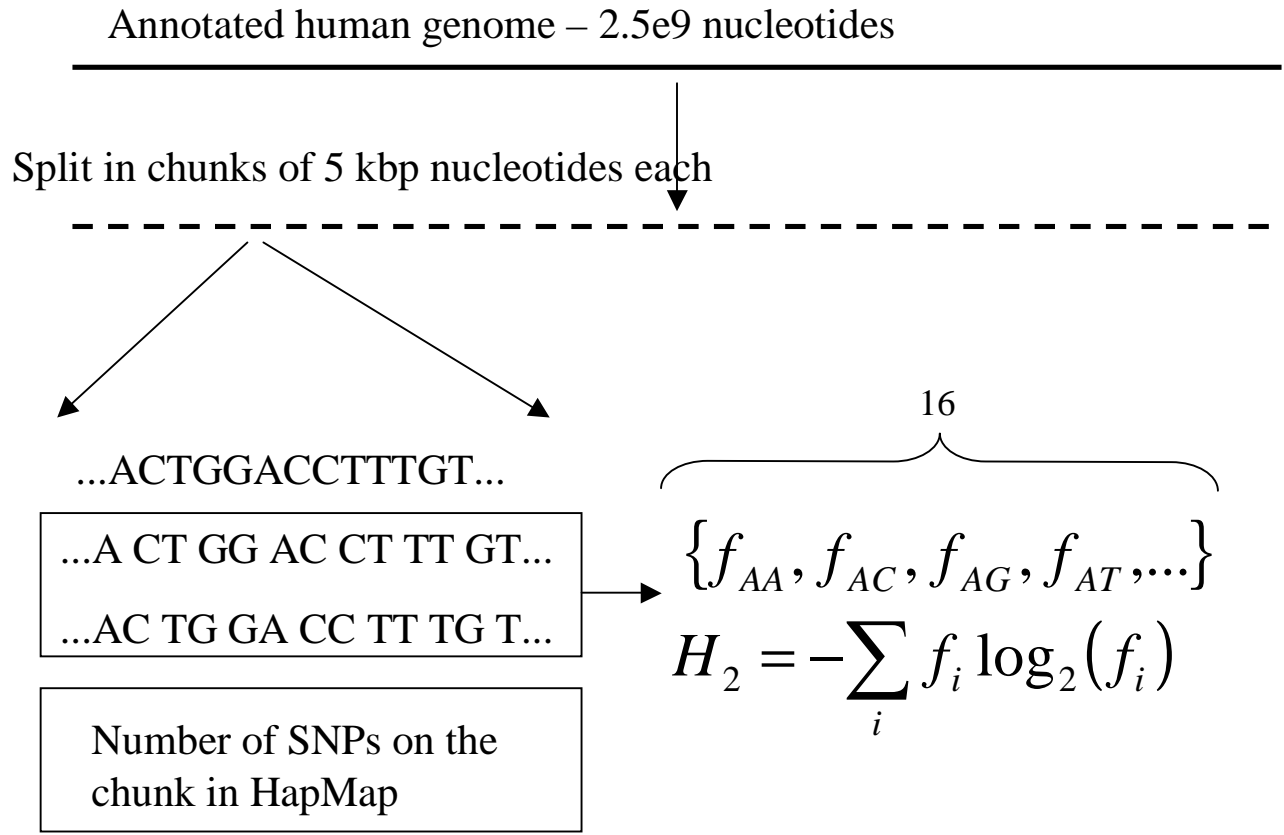

Fig. 1. Schema of data analyses. Besides the sequence entropy for each genomic piece, the number of SNP in it was extracted from HapMap, and whether it overlaps with known exons.

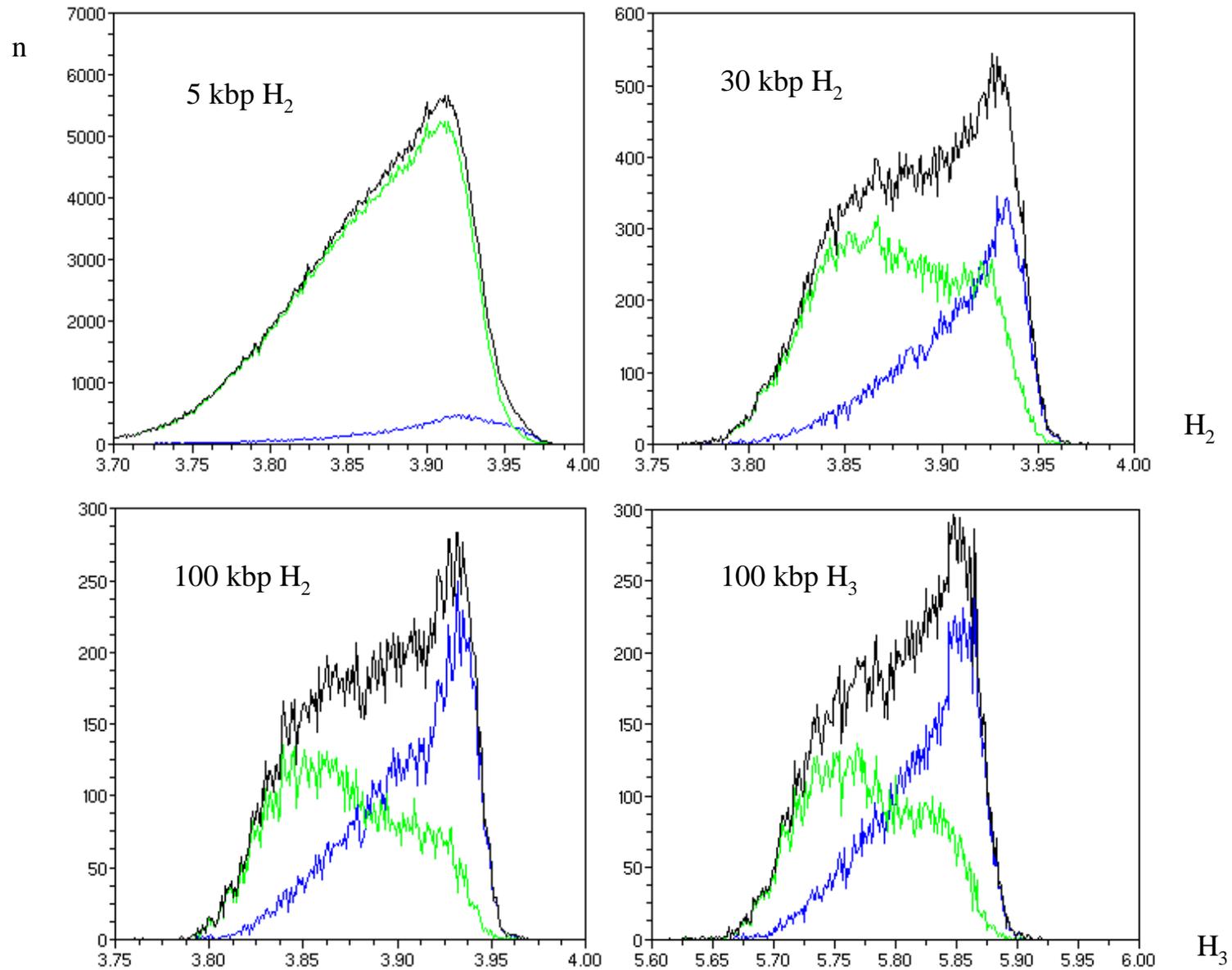

Fig. 2. Histograms of the number (n) of genomic chunks falling in the corresponding entropy bins of size 0.002. Blue line – pieces which overlap with known exons, green – containing no exons, and black – cumulative.

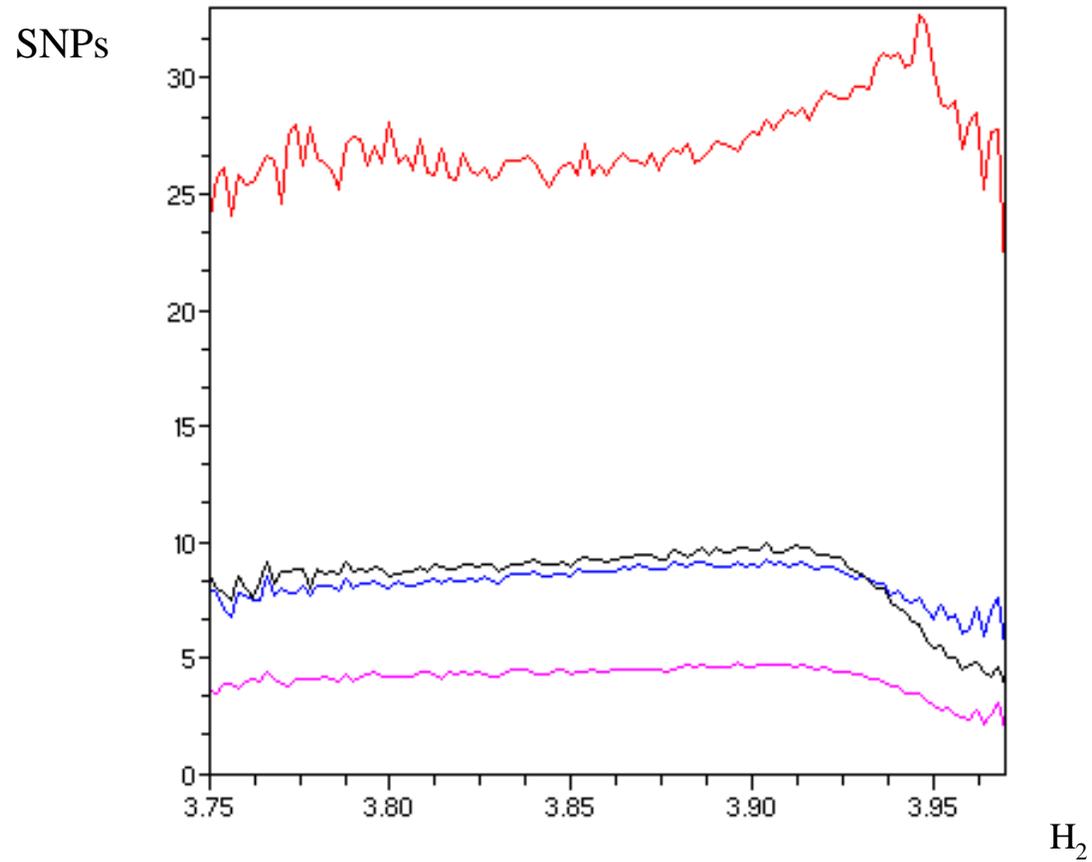

Fig. 3. Average number of SNPs for 10kbp chunks.
Black – variable HapMap SNPs, pink – non-variable HapMap SNPs
red – all dbSNP SNPs, blue - dbSNP SNPs with known average heterozygosity.

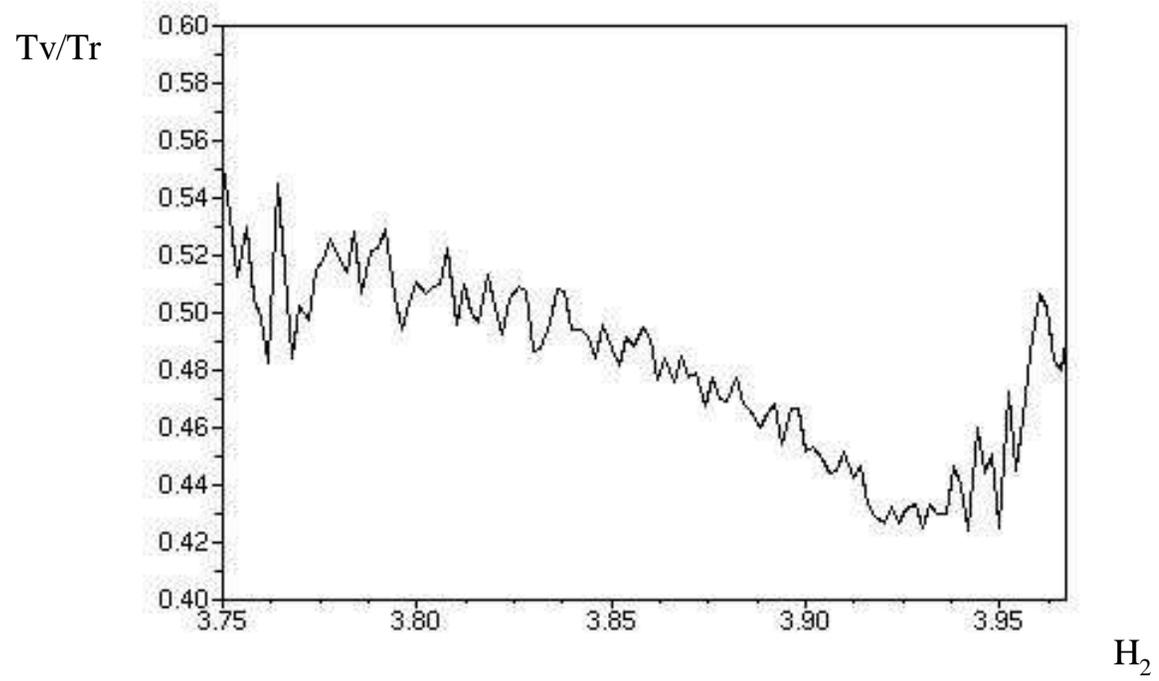

Fig. 4. Averaged transversion to transitions ratio vs. entropy. Bin size is 0.001 bits.

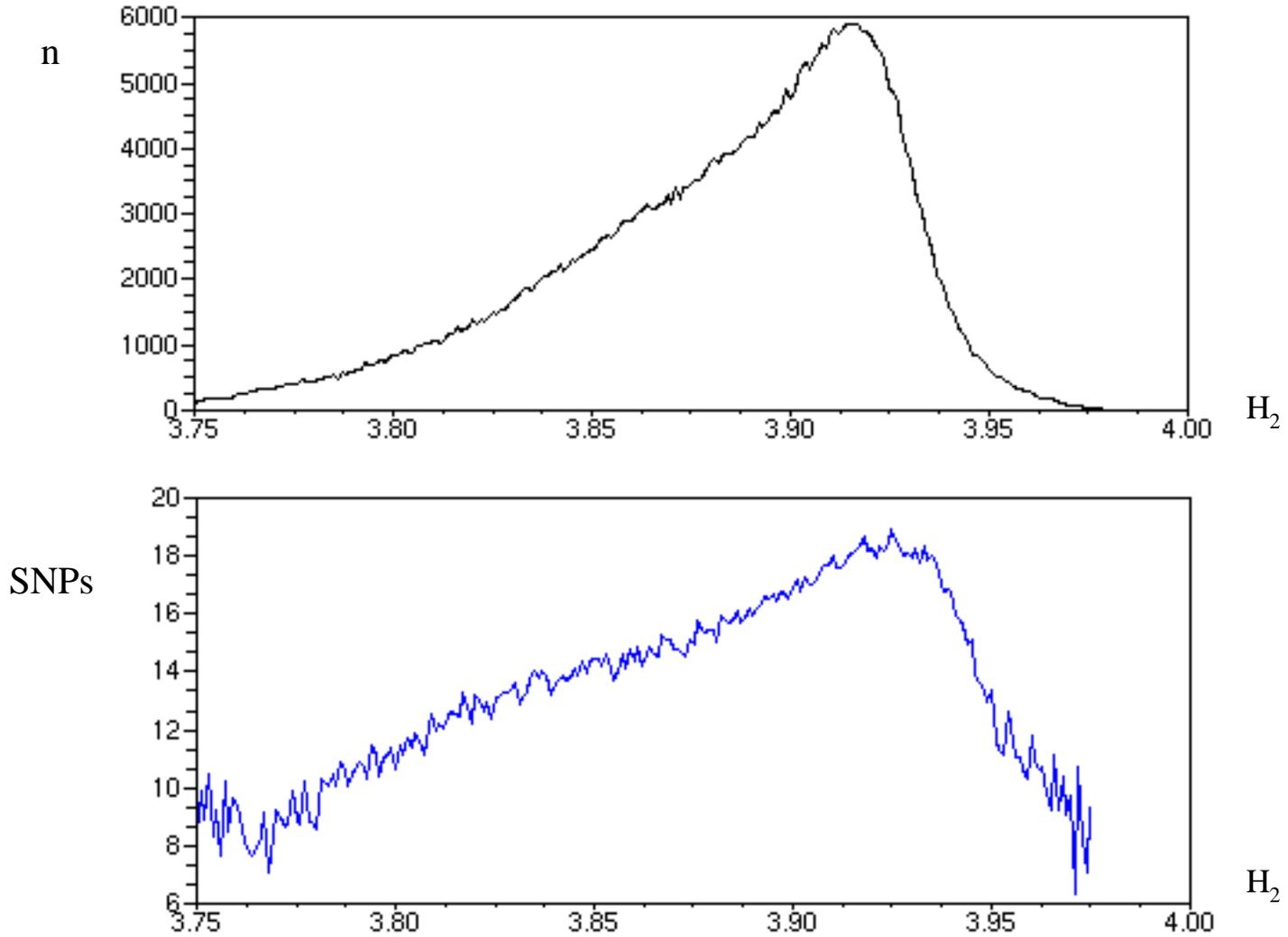

Fig. 5. Entropy distribution of 5kb chunks of mouse genome and the average SNPs number per chunk. Genomic sequence length - $2.4 \cdot 10^9$ bp, the number of SNPs - $7.5 \cdot 10^6$ mapped on the genome of C57BL/6J strain. Entropy bin size is 0.001 bits.

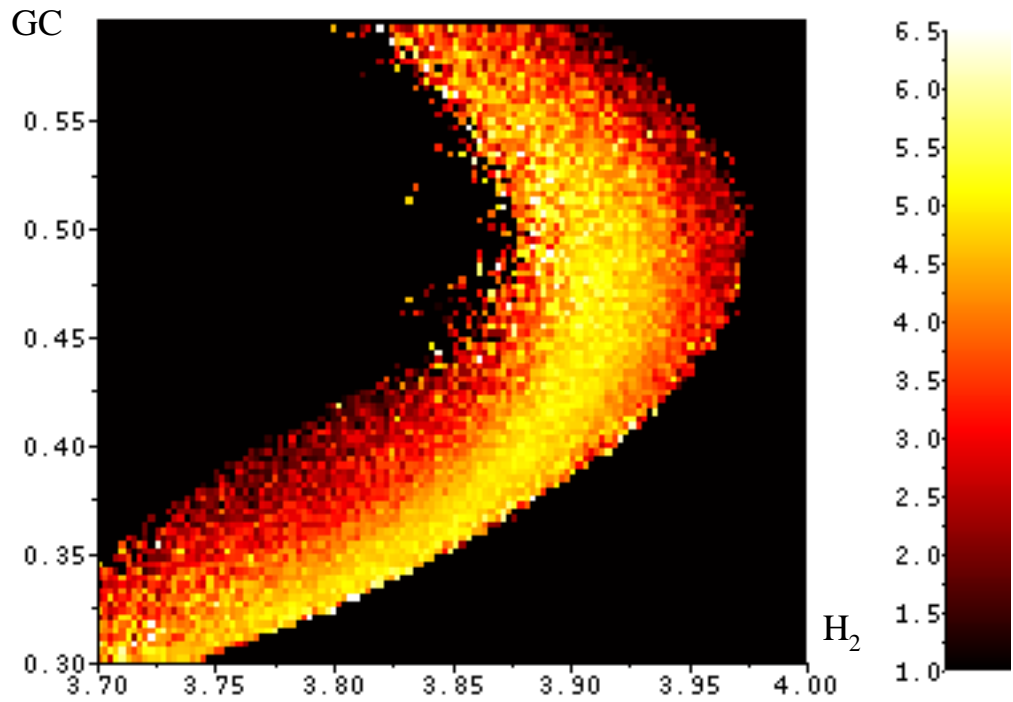

HapMap

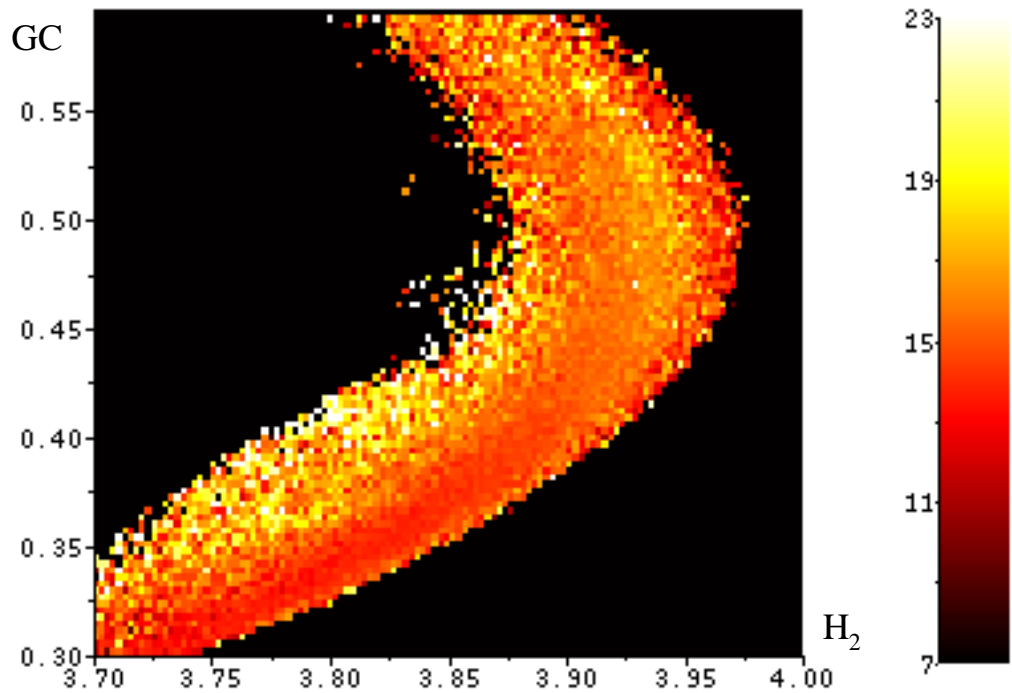

dbSNP

Fig. 6. 2D distributions of the average SNPs number per 5 kb genomic chunks falling in corresponding bin in GC-$H_2$ coordinates. Both bin sizes are 0.002.

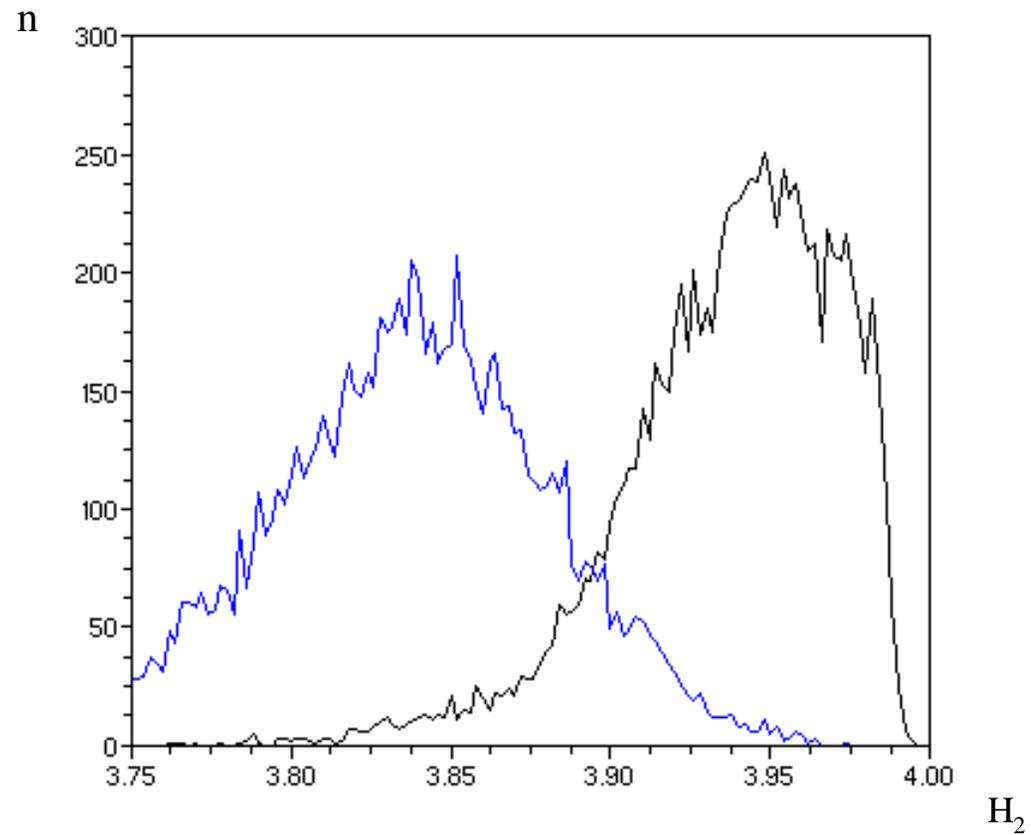

Fig. 7. Entropy distribution of 5kb chunks for *C. elegans* genome (blue) and *D. Melanogaster* genome (black).